\documentclass[12pt]{article}

\def\[{\left\lbrack}
\def\]{\right\rbrack}

\def\({\left(}
\def\){\right)}
\def\ih{\'\i}

\label{}

\begin{document}

\date{}
\title{Nonhomogeneous Cooling, Entropic Gravity and MOND Theory}
\author{Jorge Ananias Neto\footnote{e-mail: jorge@fisica.ufjf.br}\\
Departamento de F\ih sica, ICE, \\ Universidade Federal de Juiz de Fora, 36036-900,\\ Juiz de Fora, MG, Brazil }

\maketitle

\begin{abstract}
In this paper, by using the holographic principle, a modified\break equipartition theorem where we assume that below a critical temperature the energy is not equally divided on all bits, and the Unruh temperature, we derive MOND theory and a modified Friedmann equation compatible with MOND theory. Furthermore, we rederive a modified Newton's law of gravitation by employing an adequate redefinition of the numbers of bits.
\end{abstract}


\maketitle

\section{Introduction}
The use of models belonging to the different areas of physics has been  highly efficient in order to produce important theories. Gravity is a good example where pioneering works of Bekenstein\cite{Bek}, Hawking\cite{Haw} and Unruh\cite{Unr} have shown a deep connection between thermodynamics and relativity. Taking advantage of this result,
E. Verlinde\cite{Ver}  by using basically the holographic principle and the equipartition theorem has derived the usual gravitational field.  Certainly, this is an important and intriguing result. Before the work of Verlinde, some authors have also investigated gravity from a thermodynamics point of view. Among them we can cite the works of Jacobson\cite{Jac} and Padmanabhan\cite{Pad}. We should also mention that related ideas to the  Padmanabhan's works can also be found in Cantcheff\cite{MC}. In addition, Banerjee and Majhi\cite{BM} have also given a statistical interpretation of gravity. 

Having in mind a possible connection between gravity theories and solid state models, Gao\cite{Gao}, in the context of the Verlinde's formalism,  has observed that at low temperatures the equipartition 
theorem must  be corrected according to the Debye's solid model\cite{Kit}. As an important  consequence,  the gravitational force is modified. It is opportune  to mention here that the energy of a system in the Debye's model goes to zero when the temperature reaches zero. However, in the Gao's approach, the energy of the gravitational system remains constant. This behavior indicates that there is a possible difference between the Gao's procedure and the usual Debye's model. 
Li and Chang\cite{LC}, in an interesting work, have obtained the modified Newtonian dynamics (abbreviated by MOND and originally proposed by Milgrom\cite{Mond}) from the one dimensional Debye model. Also, in the Verlinde's picture,  Modesto and Randono\cite{MR} have pointed out the possibility of obtaining MOND theory from an area entropy relation and Kiselev and Timofeev\cite{KT} have related MOND theory with the collective motions of the holographic screen bits. MOND theory, in the context of cosmology, has been successful in explaining most of the observed galaxies rotation curve. As the equipartition law of energy plays an important role in the entropic gravity formalism then our purpose in this paper is to modify the equipartition theorem by adopting an alternative method to the Debye model. In the Debye's solid model the atoms are arranged in a crystal lattice structure where each can be considered as a harmonic oscillator. In our scheme, based on models of critical phenomena, during the cooling of the system, we assume that, at first, the bits are not clustered in a lattice structure. We will consider that part of the bits acquires zero energy without the system to have still reached the zero temperature. In principle, there is a critical temperature 
where this phenomenon starts to occur. When we subtract the number of bits with zero energy from the total number of bits in the equipartition formula, the MOND theory is obtained. There are several possibilities of introducing inhomogeneity in the Verlinde's formalism and our motivation in choosing a particular form, Eq.(\ref{n0}), was to derive MOND theory. In the last part of this work, we rederive  a modified form of the gravitational force, previously obtained 
 by Sheykhi\cite{AS},  by using the equipartition theorem with an appropriate redefinition of the number of bits.

\section{Entropic Gravitational Acceleration}
In this section we will briefly  show, in a quantitative manner, the steps used by Verlinde\cite{Ver} in order to derive the gravitational acceleration. We will begin by considering a spherical surface as the holographic screen with a particle of mass M positioned in its center. A holographic screen can be thought as a storage device for information. The number of bits is then assumed to be proportional to the area $A$ of the holographic screen

\begin{eqnarray}
\label{bits}
N = \frac{A}{l_p^2},
\end{eqnarray}
where $ A = 4 \pi r^2 $ and $l_p^2 = \frac{G\hbar}{c^3}$ . The term bit signifies the smallest unit of information in the holographic screen.  
The energy of the particle is 

\begin{eqnarray}
\label{en}
E = M c^2.
\end{eqnarray}
An important assumption in the Verlinde's formalism is that the total energy of the bits on the screen is given by the equipartition law of energy

\begin{eqnarray}
\label{eq}
E = \frac{1}{2}\,N k_B T.
\end{eqnarray}
It is important to mention that for black hole space-time Eq.(3) can be proved without assuming a priori the validity of this equation\cite{BM}. Assuming that the energy of the particle inside the holographic screen is equal to the equipartition law of energy

\begin{eqnarray}
\label{meq}
M c^2 = \frac{1}{2}\,N k_B T,
\end{eqnarray}
and using Eq.(\ref{bits}) together with the Unruh temperature formula

\begin{eqnarray}
\label{un}
k_B T = \frac{1}{2\pi}\, \frac{\hbar a}{c},
\end{eqnarray}
we are then able to derive the well known (absolute) gravitational acceleration formula

\begin{eqnarray}
a= \frac{G M}{r^2}.
\end{eqnarray}

\section{The MOND Theory and Modified Entropic Force}
The success of MOND theory is due to its ability, in principle as a phenomenological model, to explain the most of the rotation 
curve of galaxies. MOND theory reproduces the well known Tully-Fisher relation\cite{TF} and can also be an alternative to the dark matter model. However, it is opportune to mention that MOND theory can not explain the temperature profile of galaxy clusters and presents some trouble when confronting with cosmology. For details, see, for example, references \cite{Sk} and \cite{NP}.
Basically this theory is a modification of Newton's second law in which the force can be described as

\begin{eqnarray}
F=m\; \mu\(\frac{a}{a_0}\)\,a,
\end{eqnarray}
where $\mu(x)$ is a function with the following properties: $\mu(x)\approx 1$ for $ x > > 1$, $\mu(x) \approx x$ for $ x < < 1$
and $ a_0 $ is a constant.
There are different interpolation functions for $\mu(x)$\cite{form}. However, it is believed that the main implications caused by the MOND theory  do not depend on the specific form of these functions. Therefore, for simplicity, it is usual to assume that the variation of $\mu(x)$ between the asymptotic limits occurs abruptly at $x=1$ or $a=a_0$.

We will begin our proposal by considering that, below a critical temperature, the cooling of the holographic screen is not homogeneous. We choose that the fraction of bits with zero energy is given by the formula

\begin{eqnarray}
\label{n0}
\frac{N_0}{N} = 1-\frac{T}{T_c},
\end{eqnarray}
where $N$ is the total number of bits given by the formula (1), $N_0$ is the number of bits with zero energy and $T_c$ is the critical temperature.
For $ T \geq T_c $ we have $N_0=0$ and for $T < T_c$ the zero energy phenomenon for some bits starts to occur. 
Equation (\ref{n0}) is an usual relation of critical phenomena and second order phase transitions theory. The number of bits with energy different of zero for a given temperature $ T < T_c \,$ is

\begin{eqnarray}
\label{sub}
N-N_0 = N \frac{T}{T_c}.
\end{eqnarray}
Here, we are assuming that Eq.(\ref{bits}) is still valid below $T_c$. Whereas now the energy of the particle inside the holographic screen is equally distributed on all bits with nonzero energy and using relation (9) in the equipartition law of energy, we get

\begin{eqnarray}
\label{ebec}
M c^2 &=& \frac{1}{2} (N-N_0) k_B T \nonumber\\\nonumber\\ 
&=& \frac{1}{2}\, N \frac{T}{T_c}\; k_B T.
\end{eqnarray}
Then, combining Eqs.(1), (5) and (10), we are capable of deriving,  for $ T < T_c $, the MOND theory for Newton's law of gravitation

\begin{eqnarray}
a\,\(\frac{a}{a_0}\)=G \frac{M}{r^2},
\end{eqnarray} 
where 

\begin{eqnarray}
\label{a0}
a_0=\frac{2\pi c\, k_B  T_c}{\hbar}. 
\end{eqnarray}
Using $a_0\approx 10^{-10}\, m\,s^{-2}$ we get $T_c \approx 10^{-31} K$,  an extremely low temperature which is far from the usual temperatures observed in our real world.  Here, a brief comment on this extremely low temperature must be made. Rewriting Eq.(\ref{un}) in the form

\begin{eqnarray}
T \approx 4 \times 10^{-21} \, \frac{K}{m/s^2}\; a,
\end{eqnarray}
then, we can observe that, at first, any model that combines the Unruh temperature with MOND theory ($a_0\approx 10^{-10}\, m\, s^{-2}$) leads to the value of the critical temperature, $T_c$, mentioned above. According to the Unruh effect, where an accelerating observer will observe a black-body radiation, our critical temperature 
should be observed in an accelerated reference frame that can be the galaxies.
Therefore, below the critical temperature, $T_c$, we have obtained as a consequence of MOND theory that the usual Newton's second law is no longer valid. 

We also want to make a comment on the possibility to occur phase transitions in our model. The energy of the system, Eq.(\ref{ebec}), is a constant value ($E = M c^2$) as well as it is constant in the Debye's model as mentioned in the introduction. Then, the derivative of this expression with respect to temperature, which is the specific heat, is zero. However, so that the phase transition takes place, it is necessary that the specific heat  presents singularity at the critical temperature. Therefore, if we use the behavior of the specific heat as the indicator of phase transition, we can conclude that, strictly speaking, there is no phase transition in our particular model\cite{Huang}.

The hypothesis of nonhomogeneous cooling of bits can be justified by making use of thermostatistic arguments. Below the critical (low) temperature, the thermal bath does not ensure an exact application of the equipartition theorem for all bits, i.e., the thermalization of the system does not guarantee that all bits have the same energy. This result frequently occurs in many different physical systems. As an example, we can mention the well known coupled harmonic oscillators model\cite{chom} where the energy is not be shared between the independent normal modes of this system.

The concept of nonhomogeneous cooling of bits  allows us to imagine a curious hypothetical case that is when all bits acquire zero energy at the same time for a particular acceleration or specific critical temperature different of zero. In the language of critical phenomena, we can say that this case is similar to the first order phase transition. Then, using the abrupt energy change of the bits and Eqs.(\ref{bits}), (\ref{en}), (\ref{meq}) and (\ref{un}), we obtain that the gravitational acceleration has an interesting behavior that is not approaching to zero for very large distances but tends to a constant value. If this phenomenon occurs,
we expect that the asymptotic value of the acceleration must be very small when compared with values of our everyday world and 
certainly it is very difficult of being measured by the current experimental techniques.

\section{Modified Friedmann Equation Compatible with MOND Theory} 
 In this section we will derive a modified Friedmann equation compatible with MOND theory. We use the same steps of Cai, Cao and  Ohta\cite{CCO} who have obtained the Friedmann equations from entropic force (similar results can be found in references \cite{TP} and \cite{SY}). The only difference is that we will introduce the concept of nonhomogeneous cooling of bits in the equipartition law of energy as we did in the previous section. We will begin by considering a flat Friedmann-Robertson-Walker (FRW) universe whose metric is

\begin{eqnarray}
\label{flat}
ds^2 = - dt^2 + a^2(t) \(dr^2 + r^2 d\Omega^2\).
\end{eqnarray}
We suppose that in the FRW universe the matter source, for simplicity, is a perfect fluid with the stress-energy tensor given by

\begin{eqnarray}
T_{\mu\nu} = (\rho + \frac{p}{c^2}) u_\mu u_\nu + \frac{p}{c^2} g_{\mu\nu},
\end{eqnarray} 
where $ \rho$ is the mass density, $p$ is the pressure and $u^\mu = (1,0,0,0)$ is the four velocity of the fluid measured by a comoving observer. We choose as the holographic screen a spherical surface with the radius $ \tilde{r} = a(t) r$. Consequently, the acceleration in which will be used in the Unruh temperature formula is $ a_r = \ddot{a} r $. The mass inside the holographic screen is given by the Tolman-Komar mass relation\cite{TK}

\begin{eqnarray}
\label{tm}
M &=& 2 \int_V dV \( T_{\mu\nu} - \frac{1}{2} T g_{\mu\nu} \) u^\mu u^\nu\nonumber\\
&=&\frac{4\pi}{3}\, a^3 r^3 \, (\rho + \frac{3 p}{c^2}).
\end{eqnarray}
At this point we are ready to apply the Verlinde's formalism with the fraction bits number with zero energy,  Eq.(\ref{n0}), being used in the equipartition law of energy. Combining Eqs.(\ref{bits}), (\ref{en}), (\ref{un}), (\ref{ebec}) and (\ref{tm}), we get

\begin{eqnarray}
\label{fm}
\ddot{a} = - \[ \frac{4\pi G a_0 }{3} (\rho + \frac{3 p}{c^2})  \frac{a}{r} \]^{\frac{1}{2}}, 
\end{eqnarray}
which is the acceleration equation for the dynamical evolution of the FRW universe compatible with MOND theory. The parameter $a_0$ 
is the same as defined in  Eq.(\ref{a0}).
The minus sign in Eq.(\ref{fm}) is due to our consideration that the acceleration is caused by the matter inside the holographic screen.
Due to the algebraic form of  Eq.(\ref{fm}), at first, we can not integrate it and consequently we can not write an expression for $\dot{a}$, as usually is done for the Friedmann equations. What it is amazing in the entropic formalism is that we are deriving a modified Friedmann equation compatible with MOND theory by a thermostatistic approach and not by a modified relativistic theory\cite{Jacob} which would, in principle, be necessary to derive an exact result. 

Rewriting  Eq.(\ref{fm}) in terms of $\tilde{r}= a(t) r $, we get

\begin{eqnarray}
\label{fms}
\ddot{\tilde{r}} = - \[ \frac{4\pi G a_0 }{3} (\rho + \frac{3 p}{c^2})  \tilde{r} \]^{\frac{1}{2}}, 
\end{eqnarray}
which is the same expression obtained by Sanders\cite{San} in another context. Our derivation of the modified Friedmann equation relies on the relativistic Tolman-Komar mass formula, Eq.(\ref{tm}). This choice means that we  are adopting a particular theory of gravity. If we use another equally acceptable theory that is the Hawking-Hayward or Misner-Sharp mass formula, we can show that the results are Eqs. (\ref{fm}) and (\ref{fms}) without the pressure term. For more details of Hawking-Hayward energy and Misner-Sharp energy see references \cite{H2} and  \cite{C2}, respectively.

\section{Entropic Corrections to the Newton's Law of Gravitation}

Loop quantum gravity provides quantum corrections for the entropy-area relation of black holes. This modified formula is written as

\begin{eqnarray}
\label{slqg}
S = \frac{A}{4 l_p^2} - \beta \ln \frac{A}{4 l_p^2} + \gamma \frac{l_p^2}{A} + const.\;,
\end{eqnarray}
where $\beta$ and $\gamma$ are dimensionless constants of order unity. The values of these constants are not precisely determined  and still remain an open question in loop quantum cosmology.  Sheykhi, by utilizing the modified entropy-area relation(\ref{slqg}), has been able to derive a correction to the Newton's law of gravitation by adopting the viewpoint of Verlinde's formalism (another similar result can be found in reference \cite{LDHD}) . The result is

\begin{eqnarray}
\label{fs}
F = - \frac{ G M m}{R^2} \[ 1 - \frac{\beta}{\pi} \frac{l_p^2}{R^2} - \frac{\gamma}{4 \pi^2}\frac{l_p^4}{R^4} \].
\\ \nonumber
\end{eqnarray}
Using the FRW metric, Eq.(\ref{flat}), the definition of mass density, $\rho=M/V$, and the Tolman-Komar mass relation, Eq.(\ref{tm}),  Sheykhi has also derived from Eq.(\ref{fs})  a modified equation for the dynamical evolution of the FRW universe given by

\begin{eqnarray}
\label{frids}
\frac{\ddot{a}}{a}=-\frac{4\pi G}{3} (\rho+3 p) \[ 1 - \frac{\beta}{\pi} \frac{l_p^2}{R^2} - \frac{\gamma}{4 \pi^2}\frac{l_p^4}{R^4} \].
\end{eqnarray}
In this section  we will derive  Eqs.(\ref{fs}) and (\ref{frids}) by another procedure which consists of using the equipartition theorem together
 with  an adequate redefinition of the number of bits, $N$. Since, in the Verlinde's formalism, the definition of the number of bits, Eq.(\ref{bits}),  is, in principle, postulated then, we can adopt a new expression for the bit's number written as

\begin{eqnarray}
\label{Nc}
N^\prime= \frac{N}{1-\frac{4\beta}{N}-\frac{4\gamma}{N^2}},
\end{eqnarray}
where $N$ is given by Eq.(\ref{bits}). Hence, combining Eqs.(\ref{bits}), (\ref{en}), (\ref{eq}), (\ref{meq}), (\ref{un}) and (\ref{Nc}), we can rederive Eq.(\ref{fs}) and, consequently,  Eq.(\ref{frids}).
Here, it is interesting to note that different choices for the definition of the number of bits can result in different possible modified gravitational force expressions or different gravity theories.

\section{Conclusions}
In this work, we use the holographic principle and a modified equipartition law of energy, where we have introduced
the notion of nonhomogeneous cooling of bits, in order to derive MOND theory and a modified Friedmann equation compatible with MOND theory. The key point of our work is that, below a critical temperature, the energy of the particle inside the holographic screen is not distributed equally on all bits. This hypothesis leads to a change in the equipartition energy formula where now it is  proportional
to the square of temperature. Using the Unruh temperature we are then able to derive MOND theory.
We would like to point out that an important result of our work is the reinterpretation of MOND's parameter $a_0$, Eq.(\ref{a0}), in terms of the Planck constant and a critical temperature where below this the nonhomogeneous bits distribution of energy begins to occur. To finish, it is important to mention that our last result indicates that the quantum corrections effects of the entropy-area relation, Eq.(\ref{slqg}), can also be obtained  by taking the classical equipartition theorem in association with a suitable redefinition of the number of bits, Eq.(\ref{Nc}).

\end{document}